\documentclass[preprint,aps,12pt,preprintnumbers,eqsecnum,nofootinbib]{revtex4}
\usepackage{xcolor}
\usepackage{graphicx}
\usepackage{subfigure}
\usepackage{epstopdf} 
\usepackage{braket}
\usepackage{amssymb,amsmath,amsfonts,comment,latexsym,graphicx,epstopdf,bbold,float}
\usepackage{color}
\usepackage{fancyvrb}
\usepackage{chngcntr}
\usepackage{etoolbox}
\usepackage{bbm,epsfig}
\usepackage{array}
\usepackage{hepunits}
\usepackage{slashed}

\newcommand{\be}{\begin{equation}}
\newcommand{\ee}{\end{equation}}
\newcommand{\ba}{\begin{eqnarray}}
\newcommand{\ea}{\end{eqnarray}}

\newcommand{\grts}{\raise.3ex\hbox{$>$\kern-.75em\lower1ex\hbox{$\sim$}}}
\newcommand{\lets}{\raise.3ex\hbox{$<$\kern-.75em\lower1ex\hbox{$\sim$}}}

{\catcode`\|=\active\gdef\Braket#1{\left<\mathcode`\|"8000\let|\bravert {#1}\right>}}

\def\bravert{\egroup\,\vrule\,\bgroup}

\usepackage{subfigure}

\usepackage{color}
\usepackage{amssymb,amsmath,amsfonts}
\usepackage{epstopdf} 
\usepackage{braket}

\begin{document}
%
%
\title{\vspace*{0.5in} 
Ghosts in the Wall
\vskip 0.1in}
\author{Christopher D. Carone}\email[]{cdcaro@wm.edu}
\affiliation{High Energy Theory Group, Department of Physics,
William \& Mary, Williamsburg, VA 23187-8795}

%
%
\date{October 26, 2020}
\begin{abstract}
We consider the possibility of an axion-like particle (ALP) that is a ghost, with wrong-sign
kinetic and mass terms.  Such an ALP can arise as the partner to an ordinary particle in
theories with higher-derivative quadratic terms. We compute the photon regeneration probability in
light-shining-through-wall experiments and show that the presence of such a ghostly ALP 
can lead, in principle, to discernible effects.
\end{abstract}

\maketitle

\section{Introduction} \label{sec:intro}
The axion is the pseudo-goldstone boson of a spontaneously broken U(1) global symmetry that is
anomalous with respect to the SU(3) color gauge group of the standard model~\cite{axions}.  Since models
with axions can solve the strong CP problem~\cite{PQmech}, axion phenomenology has been the focus of
considerable attention.   Axion-like particles (ALPs) have a similar 
low-energy effective Lagrangian and can be probed via the same experimental techniques used to search for 
axions~\cite{Irastorza:2018dyq}.   Whether or not an ALP contributes to the solution of the strong CP problem, the
fact that such particles can arise so generically in extensions of the standard model makes
them of interest for phenomenological investigation.

In the case where the ALP is pseudoscalar, its coupling to photons is given by the Lagrangain
\begin{equation}
{\cal L} = -\frac{1}{4} F_{\mu\nu} F^{\mu\nu} + \frac{1}{2} \partial_\mu a \, \partial^\mu a - \frac{1}{2} m^2 a^2
+\frac{1}{4} g_{a\gamma\gamma} \,a \, \widetilde{F}_{\mu\nu} F^{\mu\nu} \,\,\, ,
\label{eq:pseudo}
\end{equation}
where, $F_{\mu\nu}$ is the photon field strength tensor, 
$\widetilde{F}^{\mu\nu} \equiv \frac{1}{2} \epsilon^{\mu\nu\alpha\beta}\, F_{\alpha\beta}$ is its dual,  and 
$g_{a\gamma\gamma}$ is a dimensionful coupling, with units of inverse mass.  In the case where the
ALP is a scalar, we may relabel $a \rightarrow s$, and replace the interaction term in Eq.~(\ref{eq:pseudo}) 
with $\frac{1}{4} \, g_{s\gamma\gamma} \,s\, F_{\mu\nu} F^{\mu\nu}$.   It has been long known that
the ALP-photon interaction may lead to the conversion of photons to ALPs in the presence of a strong magnetic 
field. While photons from an incident beam can be blocked by a barrier, the ALPs may proceed through it, and 
be converted back to photons subsequently in the presence of a magnetic field on the other side~\cite{VanBibber:1987rq}.   
Such light-shining-through-wall (LSW) experiments have provided excluded regions in the $m_A$-$g_{A\gamma\gamma}$ 
plane~\cite{Zyla:2020zbs}, for $A=a$ or $s$.

The possibility we consider here is a second ALP that appears in a higher-derivative generalization of this
effective Lagrangian, with quadratic terms
\begin{equation}
{\cal L} =  -\frac{1}{2} \hat{a} \Box \hat{a} - \frac{1}{2} m^2 \hat{a}^2 -\frac{1}{2 M^2} \, \hat{a}\, \Box^2 \hat{a} \,\,\, .
\label{eq:hdquad}
\end{equation}
In cases where the distinction between scalar and pseudoscalar does not matter, we will refer to the ALP as $a$, with a hat
to signify the higher-derivative form of the theory. In  a theory where the ALP is associated with the spontaneous breaking of a global 
symmetry via the vacuum expectation value of a complex scalar field $\phi$,  a similar higher-derivative term for $\phi$ in its 
Lagrangian leads to Eq.~(\ref{eq:hdquad}), as we will see later.  In the next section, we review how Eq.~(\ref{eq:hdquad}) can be recast 
as a theory without higher-derivative terms, but with a new field $\tilde{a}$ that has wrong sign kinetic and mass terms, {\em i.e.}, one that 
corresponds to a Lee-Wick ghost~\cite{LeeWick}.  Like the Lee-Wick Standard Model~\cite{Grinstein:2007mp}, such a theory will have a unitary 
$S$-matrix provided that (1) the Lee-Wick partner is excluded from the spectrum of asymptotic states~\cite{LeeWick} and (2) certain prescriptions 
are applied in evaluating loop diagrams~\cite{Cutkosky:1969fq}.   If one assumes that Lee-Wick partners serve as a solution to the hierarchy 
problem, then partners must be present for every particle, including in sectors that represent extensions of the Standard Model.  The 
masses of these partner particles are free parameters; no physical consideration forces a common Lee-Wick scale for all the partner particles, 
though this is often assumed in the literature as a simplifying assumption.  We will therefore take $M$ as a free parameter in our discussion.  We 
will see that the coupling of $\tilde{a}$ to photons (and in fact to anything else) will have the same form as an ordinary ALP, justifying the choice to 
call it an ALP as well.

In this paper, we investigate the effects of a light Lee-Wick ALP on LSW experiments.   The most common computation 
of the probability for regeneration of photons on the far side of a barrier first computes the probability of a photon converting to 
an ALP (and vice versa) in a magnetic field, as inferred from solutions to classical field equations (see, for example, Ref.~\cite{VanBibber:1987rq}).
This approach is not useful for taking into account the virtual Lee-Wick contribution, or its interference with the ordinary one; we instead approach the 
problem in a quantum field theory setting, maintaining the requirement that the Lee-Wick partner not appear in a Feynman diagram as an external
line.  We show that the new contribution to the scattering amplitude alters the photon regeneration probability when both the ordinary and 
Lee-Wick ALP are light.  The effect we identify is one that might be experimentally discerned.

In the next section we discuss the relevant effective theory, and in Sec.~\ref{sec:LSW} present the computation of the 
probability relevant to LSW experiments.   In the final section, we summarize our conclusions.

\section{Lee-Wick ALPs} \label{sec:LWALPS}
The procedure for converting the Lagrangian in Eq.~(\ref{eq:hdquad}) into one without the higher-derivative term, by use of an
auxiliary field, is now well known from the literature on the Lee-Wick Standard Model~\cite{Grinstein:2007mp}.  Here we give only a brief review.  

The following Lagrangian
\begin{equation}
{\cal L}_{AF} = -\frac{1}{2} \hat{a} \Box \hat{a} -\frac{1}{2} m^2 \, \hat{a}^2 - \tilde{a} \Box \hat{a} + \frac{1}{2} M^2 \tilde{a}^2 + {\cal L}_{int}(\hat{a})
\end{equation}
can be shown to be equivalent to Eq.~(\ref{eq:hdquad}) by functionally integrating over the auxiliary field $\tilde{a}$ in the generating functional for the theory.   Rewriting this in terms of the shifted field 
\begin{equation}
\hat{a} = a - \tilde{a} \,\,\, ,
\end{equation}
the Lagrangian becomes
\begin{equation}
{\cal L} = -\frac{1}{2} a \Box a + \frac{1}{2} \tilde{a} \Box \tilde{a} - \frac{1}{2} m^2 (a-\tilde{a})^2 + \frac{1}{2} M^2 \tilde{a}^2 + {\cal L}_{int} (a-\tilde{a}) \,\,\, .
\end{equation}
In the case where $m=0$, it is clear by inspection that $\tilde{a}$ corresponds to a particle of mass $M$ with wrong-sign kinetic and mass terms.  In the 
case where $m \neq 0$, there is mass mixing; the mass matrix can be diagonalized, without changing the form of the kinetic terms, via a symplectic rotation
\begin{equation}
\left(\begin{array}{c} a \\ \tilde{a} \end{array}\right) = \left(\begin{array}{cc} \cosh\theta & \sinh\theta \\ \sinh\theta & \cosh\theta \end{array}\right) 
\left(\begin{array}{c} a_0 \\ \tilde{a}_0 \end{array}\right)
\,\,\, ,
\end{equation}
leading to the form
\begin{equation}
{\cal L} =  -\frac{1}{2} a_0 \Box a_0 + \frac{1}{2} \tilde{a}_0 \Box \tilde{a}_0 - \frac{1}{2} m_0^2 \, a_0^2 + \frac{1}{2} M_0^2 \, \tilde{a}_0^2 + 
{\cal L}_{int} (e^{-\theta}[a_0-\tilde{a}_0]) \,\,\, ,
\end{equation}
where the $0$ subscript indicates a mass eigenstate.  One can either treat $(m,M)$ or $(m_0,M_0)$ as free parameters.  We will do the latter, in which case the mixing angle $\theta$ is determined by\footnote{Eq.~(\ref{eq:tt}) is equivalent to $\tanh 2\theta = - 2 \,m_0^2 \,M_0^2/(m_0^4+M_0^4)$ found in
Ref.~\cite{Carone:2009nu}.}
\begin{equation}
\tanh\theta = - \frac{m_0^2}{M_0^2} \,\,\, .
\label{eq:tt}
\end{equation}
The effect of field redefinition that we have described implies that the coupling of $a_0$ and $\tilde{a}_0$ to photons is given by
\begin{equation}
{\cal L}_{int} \supset \frac{1}{4} g_{a\gamma\gamma} \, e^{-\theta} \,(a_0 - \tilde{a}_0) \, \widetilde{F}_{\mu\nu} F^{\mu\nu}  \,\,\, ,
\label{eq:newagg}
\end{equation}
where the same substitutions described earlier may be applied in the case where the ALPs are scalar rather than pseudoscalar.  
Eq.~(\ref{eq:newagg}) will be used in the calculation of Sec.~\ref{sec:LSW}.

When the ALP arises via the spontaneous breaking of an approximate global symmetry, it is often identified via a nonlinear redefinition
of a complex scalar field that provides the symmetry-breaking vacuum expectation value $f/\sqrt{2}$:
\begin{equation}
\hat{\phi} = \frac{1}{\sqrt{2}} (\hat{\rho}+f) e^{i \hat{a}/f}  \,\,\, .
\label{eq:nldef}
\end{equation}
In this case, we assume the Lagrangian 
\begin{equation}
{\cal L} = - \hat{\phi}^* \Box \hat{\phi} - \frac{1}{M^2} \hat{\phi}^* \Box^2 \hat{\phi} + m_\phi^2\,\hat{\phi}^*\hat{\phi} - \frac{\lambda}{2} 
(\hat{\phi}^*\hat{\phi})^2 
\label{eq:u11}
\end{equation}
which includes a higher-derivative quadratic term, and provides for the spontaneous breaking of the U(1) global symmetry when $m_\phi^2 > 0$. Substituting Eq.~(\ref{eq:nldef}) into Eqs.~(\ref{eq:u11}), one obtains an effective Lagrangian for $\hat{a}$ and 
$\hat{\rho}$ of the form
\begin{equation}
{\cal L} = -\frac{1}{2} \hat{a} \Box \hat{a} -\frac{1}{2M^2} \hat{a} \Box^2 \hat{a}-\frac{1}{2} \hat{\rho} \Box \hat{\rho} -\frac{1}{2M^2} 
\hat{\rho} \Box^2 \hat{\rho} - \left[-\frac{1}{2} m_\phi^2 (\hat{\rho}+f)^2+\frac{\lambda}{8}(\hat{\rho}+f)^4\right]  + {\cal L}_{int}  \,\,\, .
\end{equation}
where ${\cal L}_{int}$ include derivative interactions involving $\hat{\rho}$ and $\hat{a}$, as well as couplings to 
standard model fields that may be arise, for example, via loops of heavy particles.   The potential for 
$\hat{\rho}$ is minimized for $f = \sqrt{2/\lambda} \, m_\phi $, and one finds that  $m_\rho^2 = 2 m_\phi^2$.  In axion models,
the scale of $f$, and hence the mass scale for $\hat{\rho}$ is extremely high; even if it were as low as a TeV in a generic ALP
model, this would place the mass scale for $\hat{\rho}$ far above the sub-meV scale of relevance to LSW experiments.  
Thus, we may work in an effective theory in which $\hat{\rho}$, or equivalently  $\rho$ and it's Lee-Wick partner
$\tilde{\rho}$, are integrated out of the theory.   We are left with
\begin{equation}
{\cal L} = -\frac{1}{2} \hat{a} \Box \hat{a} - \frac{1}{2 M^2} \, \hat{a}\, \Box^2 \hat{a}
-\frac{1}{2 M^2 f^2} \partial^\mu \hat{a} \,\partial^\nu \hat{a}\,\partial_\mu \hat{a}\,\partial_\nu \hat{a}
+\frac{1}{4} g_{a\gamma\gamma} \,\hat{a} \, \widetilde{F}_{\mu\nu} F^{\mu\nu} + \cdots \,\,\, ,
\end{equation}
where the ellipsis refer to any other induced couplings to standard model fields.  The point here is that the higher-derivative 
term that we assumed previously for $\hat{a}$ is present;  had we included a small, explicit breaking of the U(1) global 
symmetry in Eq.~(\ref{eq:u11}), we would also generate the $\hat{a}$ mass term of Eq.~(\ref{eq:hdquad}) as well.  The only new 
term is the higher-derivative interaction that is quartic in $\hat{a}$.  It is interesting to note that this term is not suppressed 
by four powers of the high scale $f$, but by $f^2 M^2$ instead, due to the fact that it originates from application of the
nonlinear field redefinition to the $1/M^2$-suppressed higher-derivative term in Eq.~(\ref{eq:u11}).  This term can lead to a 
three-body decay of $\tilde{a}$, though decays may also arise via non-derivative interactions that appear when explicit
breaking of the U(1) symmetry is taken into account.   Whether this term can have any other phenomenological consequences will 
not have any relevance to the calculation that we present in the next section.

\section{Light Shining through Walls} \label{sec:LSW}
To start, consider a toy example of the interaction between two real scalar fields $\chi$ and $\psi$ with classical 
sources, $V(x)_L$ and $V(x)_R$, via the interaction Hamiltonian
\begin{equation}
H_{\int} = \int d^4 x \left[ \chi \, \psi \, V(x)_L + \chi \, \psi \, V(x)_R \right] \,\,\, .
\end{equation}
Let us assume that $\psi$ is massless and consider one-into-one scattering of $\psi$ particles, in the presence of these sources. In addition, we assume that 
the interaction region includes a barrier that is impenetrable to $\psi$ particles; the sources $V_{L}$ ($V_{R}$) are assumed to be non-vanishing over a finite 
region to the left (right) of the barrier.   As the asymptotic states include only $\psi$ particles, and no intermediate measurement is made of the $\chi$ 
particles, the quantum mechanical scattering amplitude is given by
\begin{equation}
i {\cal M} = - \int d^4 x \int d^4 x' D_F(x-x') V(x)_L V(x')_R \langle p' | \psi(x) \psi(x') | p \rangle  \,\,\, ,
\end{equation}
where $D_F(x-x')$ is the Feynman propagator for the $\chi$ field.   After a series of elementary manipulations, this can be reduced to the 
momentum space expression
\begin{equation}
i {\cal M} = \int \frac{d^4 q}{(2 \pi)^4} \frac{i}{q^2-m_\chi^2+i\epsilon} \left[ i \widetilde{V}(p'-q)_L \right] \left[ i \widetilde{V}(q-p)_R\right] \,\,\, ,
\label{eq:toyamp}
\end{equation}
where $\widetilde{V}$ is the Fourier transform of a given source (following the conventions of Peskin and Schroeder~\cite{Peskin:1995ev}) 
and $p$ ($p'$) represents the four-momentum of the incoming (outgoing) $\psi$ particle.  The arguments of $\widetilde{V}$ represent the momenta 
flowing out of each source; conservation of momentum at the vertices leaves one momentum, $q$, unconstrained, which is integrated over in the amplitude.

The analogous approach may be applied to the ALP-photon vertex that we encountered previously, with combinatoric factors included to take
into account that there are two ways to choose which field strength tensor in each vertex is treated as the classical source.  Assume for the moment 
that the Lee-Wick ALP is absent (which is useful for comparing our result to the one obtained by the classical approach in Ref~\cite{VanBibber:1987rq}).  
The amplitude that is analogous to Eq.~(\ref{eq:toyamp}) is
\begin{equation}
i {\cal M} = \frac{g_{a\gamma\gamma}^2}{4} \, \epsilon^{\mu\nu\alpha\beta} \epsilon^{\rho\sigma\kappa\eta} \epsilon_\mu(p)  \epsilon_\rho(p')^*p_\nu p'_\sigma
\int \frac{d^4 q}{(2\pi)^4} \frac{i}{q^2-m_0^2+i\epsilon} \left[i \widetilde{F}^{cl}_{\alpha\beta}(q-p)_L\right] \left[i \widetilde{F}^{cl}_{\kappa\eta}(p'-q)_R\right]  \,\, ,
\label{eq:im1}
\end{equation}
where the $\epsilon(p)$ and $\epsilon(p')^*$ four-vectors encode the photon polarizations.  The tildes in Eq.~(\ref{eq:im1}) indicates a Fourier 
transform, not a dual tensor;  the latter is taken into account via the Levi-Civita symbols in the prefactor. We now work in the laboratory frame, where 
we assume a fixed magnetic field $\vec{B}$ in the $z$ direction, and consider polarized photons moving in the $x$-direction, with their electric fields 
aligned with the applied $B$-field, namely
\begin{equation}
F^{\mu\nu} = \left(\begin{array}{cccc} 0 & 0 & 0 & 0 \\ 0 & 0 & -B_z & 0 \\ 0 & B_z & 0 & 0 \\0 & 0 & 0 & 0 \end{array}\right)  \,\,\, , \,\,\,
p^\mu = \left(\begin{array}{c} \omega \\ \omega \\ 0 \\ 0 \end{array}\right) \,\,\, , \,\,\,
\epsilon^\mu = \left(\begin{array}{c} 0 \\ 0 \\ 0 \\ 1 \end{array}\right) \,\,\, .
\end{equation} 
To proceed, it is simplest to substitute these choices into Eq.~(\ref{eq:im1}), square the amplitude, and sum over the final photon polarization states, yielding
\begin{equation}
| {\cal M}|^2 = g_{a\gamma\gamma}^4 \omega^2 ({p'}_0^2-{p'}_3^2) | I_0(p',p)|^2 \,\,\, ,
\end{equation}
where
\begin{equation}
I_0(p',p)=\int \frac{d^4 q}{(2 \pi)^4} \frac{\widetilde{B}^{(L)}_z(q-p) \widetilde{B}^{(R)}_z(p'-q)}{q^2-m_0^2+i\epsilon} \,\,\, .
\label{eq:i0}
\end{equation}
We assume that the magnetic field is constant, with magnitude $B_0$, and significantly greater in spatial extent than the incoming photon beam 
in the $y$ and $z$ directions. In the $x$-direction, the magnetic field is constant over the finite interval $-L/2 \leq x \leq L/2$ (to the left of the wall) 
and $ L/2 \leq x \leq 3 L/2$ (to the right of the wall) and vanishing otherwise; the wall is assumed to be thin and located at $x=L/2$.  With these choices,
the Fourier transform of the magnetic field in the $x$ direction 
\begin{equation}
\int dx \, B^{(L)}_z(x)\, e^{-i q x} = B_0\, L \, \left[\frac{\sin(q L/2)}{qL/2}\right] \equiv B_0\, L\, F(q) \,\,\, ,
\label{eq:leftFT}
\end{equation}
where the Fourier transforms in the $0$, $2$ and $3$ directions yield delta functions.  Here and henceforth, we use $q$ to represent one-dimensional
momentum in the $x$-direction, rather than a four-momentum.  For the source on the right of the wall,
\begin{equation}
\int dx \, B^{(R)}_z(x)\, e^{-i q x} =\int dx \, B^{(L)}_z(x-L)\, e^{-i k q} = B_0 \, L \, e^{-i\, q\, L} F(q) \,\,\, .
\label{eq:rightFT}
\end{equation}
Combining Eqs.~(\ref{eq:i0}), (\ref{eq:leftFT}) and (\ref{eq:rightFT}), one may write the squared amplitude as
\begin{equation}
|{\cal M}|^2 = g_{a\gamma\gamma}^4 \, \omega^4 \, B_0^4 \,L^4 \left[(2 \pi)^3 \delta({p'}^0-\omega)\delta({p'}^2)\delta({p'}^3)
\right]^2 |I(\omega)|^2
\end{equation}
where $I(\omega)$ is the one-dimensional integral
\begin{equation}
I(\omega) = \int \frac{dq}{2\pi} \, \frac{F(q-\omega)^2 \, e^{i (q-\omega) L}}{q^2 - (\omega^2-m_0^2)- i \epsilon}  \,\,\, .
\end{equation}
We note that the quantity in brackets can be written as $[(2 \pi)^3 \delta^{(3)}(0)] \delta({p'}^0-\omega)\delta({p'}^2)\delta({p'}^3)$, while
the scattering probability is determined by
\begin{equation}
dP = \frac{1}{\langle i | i \rangle} \, \frac{d^3 p'}{(2 \pi)^3} \, \frac{1}{2 \omega'} \, | {\cal M} |^2 \equiv
\frac{1}{2 \omega [(2 \pi)^3 \delta^{(3)}(0)]} \, \frac{d^3 p'}{(2 \pi)^3} \, \frac{1}{2 \omega'} \, | {\cal M} |^2  \,\,\, .
\end{equation}
The prefactor takes into account the normalization of the incoming scattering states.  The divergent factor cancel (or, 
more rigorously, can be cancelled after regulating the delta functions by making the four-volume of the universe finite), while the 
remaining three delta functions are eliminated by the phase space integration.  We are left with the scattering probability
\begin{equation}
P = \frac{1}{4} g_{a\gamma\gamma}^4 B_0^4 \,L^4\,\omega^2\, |I(\omega)|^2 \,\,\, ,
\end{equation}
with $I(\omega)$ as given previously.   We now evaluate that integral: For convenience, we first shift the integration variable
\begin{equation}
I(\omega) = \int \frac{dq}{2\pi} \, \frac{F(q)^2 \, e^{i q L}}{(q+\omega)^2 - (\omega^2-m_0^2)- i \epsilon}  \,\,\,,
\end{equation}
and note that $F(q)^2 e^{i q L}$ has no singularities for any finite complex q, and depends only on non-negative powers of $e^{i q L}$.  Hence, we can evaluate $q$ the integral along the real axis of the complex $q$-plane by closing the contour at infinity via a semi-circle in the upper half-plane.  Defining
\begin{equation}
k_a = \sqrt{\omega^2-m_0^2} \,\,\, ,
\end{equation}
the contour just described encloses a pole at $q_*=k_a - \omega$.   By the residue theorem, one thus finds
\begin{equation}
P = \frac{1}{16} g_{a\gamma\gamma}^4 B_0^4 \,L^4\,\left(\frac{\omega}{k_a}\right)^2 F(q_*)^4  \,\,\, .
\label{eq:pcomp}
\end{equation}
This agrees with the expression quoted in Ref.~\cite{VanBibber:1987rq}, up to a different sign convention for 
the momentum $q_*$.  

It is interesting to note that the approach of Ref.~\cite{VanBibber:1987rq} involves squaring a $\gamma \leftrightarrow a$ transition probability that 
is inferred from a classical field amplitude.  Using quantum field theory, if one computes the probability of one-into-one $\gamma \rightarrow a$ 
 scattering off a single source (say the one on the left of the barrier), one finds that the square of this probability also agrees with Eq.~(\ref{eq:pcomp}). Though that may provide convenient short cut to obtaining the final answer, such a calculation is not the appropriate one:  It corresponds to a different physical situation, one in which a measurement is made of the ALP directly, an assumption that cannot apply to the second, Lee-Wick ALP.  Our approach does not run into this difficulty.

It is straightforward to include the effect of the Lee-Wick ALP in our approach, by modifying
\begin{equation}
\Big| \frac{i \omega}{2 k_a} F(q_*)^2 e^{i q_* L} \Big|^2 \longrightarrow 
\Big| \frac{i \omega}{2 k_a} F(q_*)^2 e^{i q_* L} -  \frac{i \omega}{2 \tilde{k}_a} F(\tilde{q}_*)^2 e^{i \tilde{q}_* L} \Big|^2 \,\,\, ,
\label{eq:mod}
\end{equation}
where $\tilde{q}_*$ and $\tilde{k}_a$ are defined in the same way as $q_*$ and $k_a$, with $m_0$ replaced by $M_0$; the sign
difference between terms reflects the difference in the sign of the propagators for the ordinary and Lee-Wick ALP, as well as a sign flip 
at each vertex.  After including the appropriate dependence on mixing angle in the prefactor, the conversion probability becomes
\begin{equation}
P = \frac{1}{16} g_{a\gamma\gamma}^4 B_0^4 \,L^4 \, e^{-4 \theta} G(\omega,\, m_0,\, M_0,\, L) \,\,\, ,
\label{eq:prob}
\end{equation}
where
\begin{eqnarray}
G(\omega,\, m_0,\, M_0,\, L) & = &
\left[ \left(\frac{\omega}{k_a}\right)^2 F(q_*)^4 +\left(\frac{\omega}{\tilde{k}_a}\right)^2 F(\tilde{q}_*)^4  \right. \nonumber \\
&& \left. -2 \left(\frac{\omega}{k_a}\right) \left(\frac{\omega}{\tilde{k}_a}\right) F(q_*)^2 F(\tilde{q}_*)^2 \cos[(q_* - \tilde{q}_*)L]
\right] \,\,\, .
\label{eq:G}
\end{eqnarray}

\begin{figure}[t]
\centering
\subfigure{}
 \includegraphics[width=.4\textwidth]{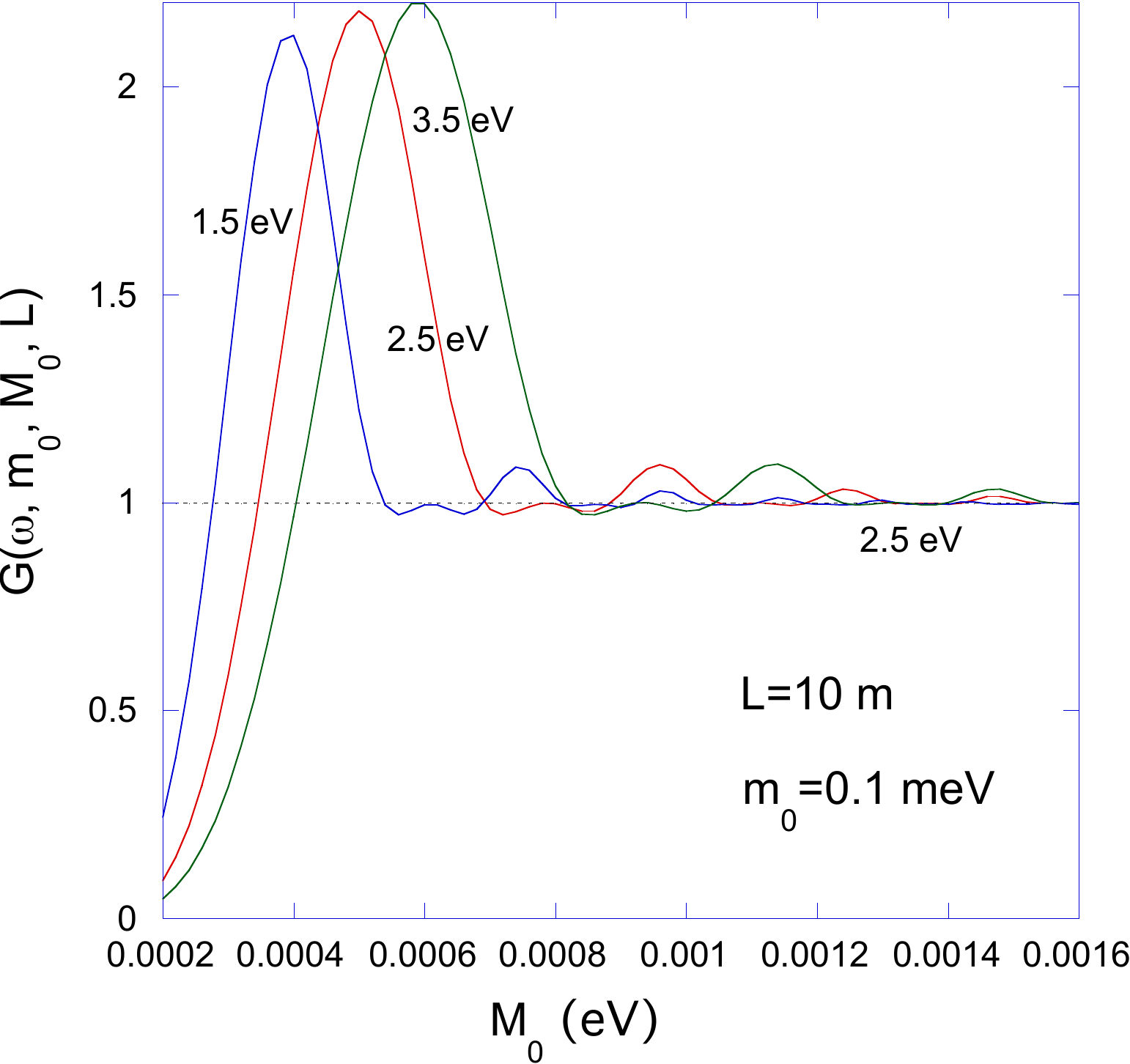}
\subfigure{}
 \includegraphics[width=.4\textwidth]{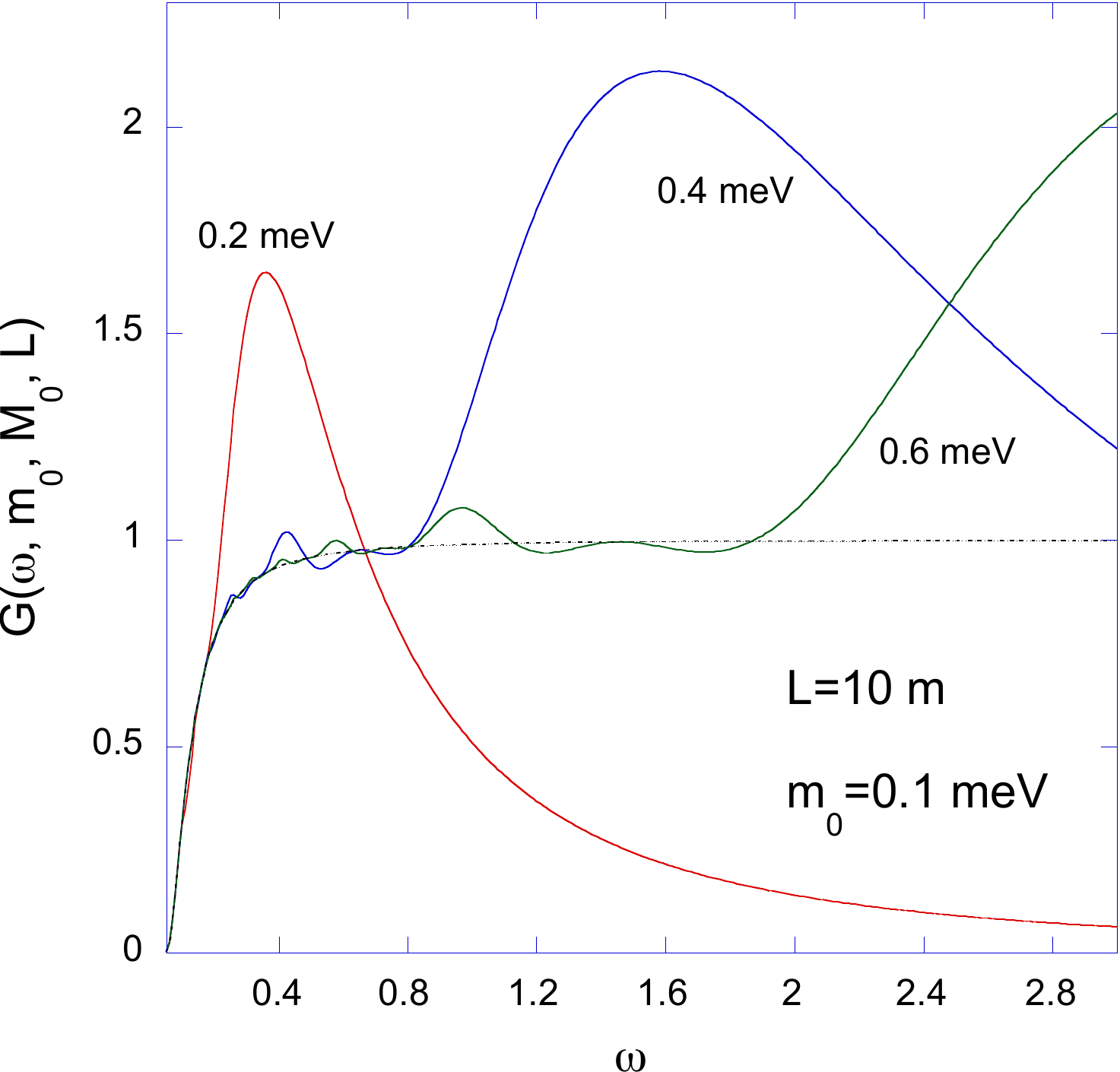}
\caption{The function $G$ defined in the expression for the photon regeneration probability, Eq.~(\ref{eq:prob}), for fixed $L$ and
$m_0$.  The curves in the left panel are labeled by photon energy $\omega$; in the right panel, by the Lee-Wick ALP mass $M_0$.
\label{fig:fig1}}
\end{figure}
 
We plot the function $G(\omega,\, m_0,\, M_0,\, L)$ in Fig.~\ref{fig:fig1}, as a function of $M_0$ (left) and photon energy $\omega$ in
the left and right panels, respectively, with other parameters held fixed.    For the purpose of illustration, we have fixed $L=10$~m and 
$m_0 = 0.1$~meV in each of these subfigures.   We note that $L=10$~m and $\omega=2.5$~eV were the benchmark values assumed in 
Ref.~\cite{VanBibber:1987rq}; moreover, current exclusion regions for axions from LSW experiments like OSQAR~\cite{Ballou:2015cka} 
go up to about $1$~meV, which motivates the $M_0$ range in the left panel of Fig.~\ref{fig:fig1}. The line that is at (or asymptotes to) $G=1$ 
corresponds to the case where $M_0=\infty$, {\em i.e.}, where only the normal ALP is present, and is provided as a point of comparison.  Note that
the separate factor of $e^{-4 \theta}$ in Eq.~(\ref{eq:prob})  can be computed from Eq.~(\ref{eq:tt}) and is of ${\cal O}(1)$ for values of $m_0$ and 
$M_0$ relevant in these figures; for a fixed choice of $m_0$ and $M_0$, this factor can be absorbed into a redefinition of the coupling $g_{a\gamma\gamma}$. 
These figures illustrate that the presence of the Lee-Wick ALP can lead to non-negligible differences in the photon regeneration probability compared
to ALP models without the higher-derivative quadratic terms.  The right panel of the figure makes clear that this difference might be probed 
experimentally by performing LSW experiments at different photon energies, where the energy dependence of the conversion probability could 
become apparent.

A number of comments are in order:  (1) One might object that the sensitivities of current LSW experiments are superseded by other 
bounds on light particles, particularly from searches for solar axions (see the bound from the CAST experiment~\cite{Anastassopoulos:2017ftl} 
in the exclusion plot in Ref.~\cite{Zyla:2020zbs}).   While this is true, the present result is of interest due to the substantial room for 
improvement in the sensitivity anticipated for future LSW experiments, corresponding to values of $g_{A\gamma\gamma}$ that are more than 
four orders of magnitudes smaller than those accessible by LSW experiments today.  This is illustrated by the projected exclusion region for the 
proposed ALPS-II~\cite{Bahre:2013ywa}, ALPS-III~\cite{lindner} and STAX2~\cite{Capparelli:2015mxa} experiments in Fig.~6 of Ref.~\cite{Irastorza:2018dyq}.   The basic qualitative conclusion of the present work, that it may be useful to look at the dependence of the conversion probability on photon energy to search for effects like those shown in the right panel of Fig.~\ref{fig:fig1}, is potentially of value.  (2) Those who have worked with Lee-Wick theories in other contexts may have anticipated a suppression of the squared amplitude due to the cancellation between the ordinary and the 
Lee-Wick particle propagators (the effect that leads to better convergence of loop diagrams in the Lee-Wick standard model).  However, this is 
not the case here, since each term in the amplitude is a finite integral in $q$ that is dominated by $q_*$, as is clear from Eq.~(\ref{eq:G}), and 
$q_* \ll m_0$.  (Note that when $m_0 \ll \omega$, as we have been assuming, $q_* \approx m_0^2/(2 \omega) \ll m_0$.)  Hence, the ultraviolet behavior of the integrand is not determining the final result. (3)  We presented the form of the photon-ALP vertex in the case where the ALP is scalar rather than pseudoscalar
in Sec.~\ref{sec:intro}.  In this case, the amplitude given in Eq.~(\ref{eq:im1}) would be modified by the replacement
\begin{equation}
\epsilon^{\mu\nu\alpha\beta} \epsilon^{\rho\sigma\kappa\eta} \longrightarrow (g^{\mu\alpha}g^{\nu\beta} - g^{\nu\alpha}g^{\mu\beta})(g^{\rho\kappa}g^{\sigma\eta} - g^{\sigma\kappa}g^{\rho\eta}) \,\,\, .
\end{equation}
The subsequent derivation is the same as the one described in the case of a pseudoscalar ALP, and the final result turns out to be 
identical to the one given in Eqs.~(\ref{eq:prob}) and (\ref{eq:G}), with only the notational replacement 
$g_{a\gamma\gamma} \rightarrow  g_{s\gamma\gamma}$. (4) The Lee-Wick ALP could have a width $\Gamma$ that is larger in magnitude 
than the ordinary ALP, depending on the model.  When a width is included, the exact Lee-Wick propagator has a pair of complex conjugate
poles and a cut.  In the narrow width approximation, one of these poles and the cut piece cancel, so that a single pole remains, with $M^2$
replaced by  $M^2 + i M | \Gamma|$; this expression reflects the fact that Lee-Wick particles have $\Gamma < 0$.    Notice that the width 
moves location of propagator poles in a direction opposite that given by the $i \epsilon$ prescription.   The Lee-Wick prescription~\cite{LeeWick}  
requires that an integration contour that is initially on the real axis and given by the Feynman prescription is deformed as the width is turned 
on, so that the poles remain in the same relative location, either above or below the contour, without crossing it.   The end result in the 
present calculation would be to replace $M^2$ by $M^2 + i M | \Gamma|$ in Eq.~(\ref{eq:mod}), and modify Eqs.~(\ref{eq:G}) accordingly.  
We have done this exercise and checked that including a width of order $0.01 M_0 - 0.1 M_0$ in our previous calculation does not change the results shown in Fig.~\ref{fig:fig1} qualitatively.

\section{Conclusions} \label{sec:conc}
We have considered the consequences of higher-derivative quadratic terms in models with Axion-Like Particles (ALPs).  Such higher-derivative
terms have been of interest in extensions of the standard model intended to address the hierarchy problem~\cite{Grinstein:2007mp}, as well as in 
attempts to formulate renormalizable theories of quantum gravity~\cite{Modesto:2016ofr}.   Their generic presence is well motivated and the mass 
scales associated with these terms are free parameters.  Higher-derivative quadratic terms in  ALP models give rise to a second ALP, a Lee-Wick 
partner~\cite{LeeWick}.  Their appropriate treatment in quantum field theory is well known~\cite{LeeWick,Cutkosky:1969fq}, and requires that they 
be excluded from the spectrum of asymptotic scattering states.  Nevertheless, if  the Lee-Wick ALP is not decoupled from the relevant low-energy 
effective theory, it can have observable consequences.  We have shown this in the context of Light-Shining-through-Walls (LSW) experiments, 
where the ordinary and Lee-Wick ALP have similar interaction vertices with photons (up to a sign), and can both contribute to the photon 
regeneration amplitude.   Our result shows a distinctive dependence of the regeneration probability on incident photon energy.  With substantial 
improvements in the sensitivity of LSW experiments anticipated~\cite{Bahre:2013ywa,lindner,Capparelli:2015mxa}, such an effect might be 
relevant experimentally in distinguishing the possibiity considered here from more conventional ALP scenarios.   

\begin{acknowledgments}  
We are grateful to Josh Erlich and Marc Sher for useful comments.  We thank the NSF for support under Grant PHY-1819575.  
\end{acknowledgments}



\begin{thebibliography}{99}

\bibitem{axions}
S.~Weinberg,
``A New Light Boson?,''
Phys. Rev. Lett. \textbf{40}, 223-226 (1978);
F.~Wilczek,
``Problem of Strong  $P$  and  $T$  Invariance in the Presence of Instantons,''
Phys. Rev. Lett. \textbf{40}, 279-282 (1978).

\bibitem{PQmech}
R.~D.~Peccei and H.~R.~Quinn,
``CP Conservation in the Presence of Instantons,''
Phys. Rev. Lett. \textbf{38}, 1440-1443 (1977);
R.~D.~Peccei and H.~R.~Quinn,
``Constraints Imposed by CP Conservation in the Presence of Instantons,''
Phys. Rev. D \textbf{16}, 1791-1797 (1977).

\bibitem{Irastorza:2018dyq}
I.~G.~Irastorza and J.~Redondo,
``New experimental approaches in the search for axion-like particles,''
Prog. Part. Nucl. Phys. \textbf{102}, 89-159 (2018)
[arXiv:1801.08127 [hep-ph]].

\bibitem{VanBibber:1987rq}
K.~Van Bibber, N.~R.~Dagdeviren, S.~E.~Koonin, A.~Kerman and H.~N.~Nelson,
``Proposed experiment to produce and detect light pseudoscalars,''
Phys. Rev. Lett. \textbf{59}, 759-762 (1987).

\bibitem{Zyla:2020zbs}
See {\em Axions and Other Similar Particles} in 
P.~A.~Zyla \textit{et al.} [Particle Data Group],
``Review of Particle Physics,''
PTEP \textbf{2020}, no.8, 083C01 (2020).

\bibitem{LeeWick}
T.~D.~Lee and G.~C.~Wick,
``Negative Metric and the Unitarity of the S Matrix,''
Nucl. Phys. B \textbf{9}, 209-243 (1969);
``Finite Theory of Quantum Electrodynamics,''
Phys. Rev. D \textbf{2}, 1033-1048 (1970).

\bibitem{Grinstein:2007mp}
B.~Grinstein, D.~O'Connell and M.~B.~Wise,
``The Lee-Wick standard model,''
Phys. Rev. D \textbf{77}, 025012 (2008)
[arXiv:0704.1845 [hep-ph]].

\bibitem{Cutkosky:1969fq}
R.~E.~Cutkosky, P.~V.~Landshoff, D.~I.~Olive and J.~C.~Polkinghorne,
``A non-analytic S matrix,''
Nucl. Phys. B \textbf{12}, 281-300 (1969).

\bibitem{Carone:2009nu}
C.~D.~Carone and R.~Primulando,
``Constraints on the Lee-Wick Higgs Sector,''
Phys. Rev. D \textbf{80}, 055020 (2009)
[arXiv:0908.0342 [hep-ph]].

\bibitem{Peskin:1995ev}
M.~E.~Peskin and D.~V.~Schroeder,
``An Introduction to quantum field theory,''
Westview Press, Boulder, CO, 1995.

\bibitem{Ballou:2015cka}
R.~Ballou \textit{et al.} [OSQAR],
``New exclusion limits on scalar and pseudoscalar axionlike particles from light shining through a wall,''
Phys. Rev. D \textbf{92}, no.9, 092002 (2015)
[arXiv:1506.08082 [hep-ex]].

\bibitem{Anastassopoulos:2017ftl}
V.~Anastassopoulos \textit{et al.} [CAST],
``New CAST Limit on the Axion-Photon Interaction,''
Nature Phys. \textbf{13}, 584-590 (2017)
[arXiv:1705.02290 [hep-ex]].

\bibitem{Bahre:2013ywa}
R.~B\"ahre, B.~D\"obrich, J.~Dreyling-Eschweiler, S.~Ghazaryan, R.~Hodajerdi, D.~Horns, F.~Januschek, E.~A.~Knabbe, A.~Lindner and D.~Notz, \textit{et al.}
``Any light particle search II \textemdash{}Technical Design Report,''
JINST \textbf{8}, T09001 (2013)
[arXiv:1302.5647 [physics.ins-det]].

\bibitem{lindner}
A. Lindner, Physics Beyond Colliders Annual Workshop, CERN, November 2017, https://indico.cern.ch/event/644287/.

\bibitem{Capparelli:2015mxa}
L.~Capparelli, G.~Cavoto, J.~Ferretti, F.~Giazotto, A.~D.~Polosa and P.~Spagnolo,
``Axion-like particle searches with sub-THz photons,''
Phys. Dark Univ. \textbf{12}, 37-44 (2016)
[arXiv:1510.06892 [hep-ph]].

\bibitem{Modesto:2016ofr}
L.~Modesto,
``Super-renormalizable or finite Lee\textendash{}Wick quantum gravity,''
Nucl. Phys. B \textbf{909}, 584-606 (2016)
[arXiv:1602.02421 [hep-th]].

\end{thebibliography}
\end{document}